\documentstyle[ltwol,epsfig]{article}

\def\be{\begin{equation}}
\def\ee{\end{equation}}
\def\bea{\begin{eqnarray}}
\def\eea{\end{eqnarray}}

\begin{document}

\title{DETECTORS AS A FUNCTION OF LUMINOSITY AT $e^+ e^-$ MACHINES}

\author{G. Eigen}

\address{Dept. of Physics, University of Bergen, Allegaten 55,
N-5007 Bergen, Norway \\E-mail: eigen@asfys2.fi.uib.no} 


\twocolumn[\maketitle\abstracts{The performance of silicon-strip 
vertex detectors, drift chambers, CsI(Tl) electromagnetic calorimeters, 
and level-1 triggers used in present multipurpose detectors like BABAR
is discussed in the context of operations 
in a high-luminosity environment.}] 

\section{Introduction}

The successful start of the asymmetric B-factories last year has triggered 
discussions on physics accessible at luminosities of 
${\cal L} = 10^{35} - 10^{36} \rm cm^{-2} s^{-1}$.\cite{phys} 
Since such luminosities are technically feasible, 
questions have been raised, how subsystems of present multipurpose detectors
like BABAR would perform in such a high-luminosity environment.
This article presents a look at the impact of machine-related backgrounds on
silicon-vertex detectors, drift chambers, CsI(Tl) electromagnetic 
calorimeters and level-1 (L1) trigger rates. Results are based on a BABAR 
study,\cite{babar} which was facilitated to cope with the planned PEP~II 
luminosity upgrade relying on the experience with the response of
BABAR to machine-related backgrounds in the first 
year of running. The data were parameterized in terms of first or 
second-order polynomials of beam currents and luminosity. For many
measurements a linear dependence was already sufficient.  
In this article extrapolations are considered for three luminosity points: 
${\cal L} =0.65 \times 10^{34} \rm \ cm^{-2} s^{-1}$, $1.5 \times 
10^{34}\rm \ cm^{-2} s^{-1}$, and  $5 \times 10^{34}\rm \ cm^{-2} s^{-1}$.
Expectations for the first two luminosity
points are taken from the BABAR study,\cite{babar}, while those for
the third luminosity point represent my personal view. Without new data, I
consider extrapolations to luminosities beyond  $ {\cal L} > 5 \times 
10^{34}\rm \ cm^{-2} s^{-1}$ unrealistic, because of large 
extrapolation uncertainties and additional backgrounds such as luminosity 
lifetime that become important at very high luminosities.\cite{MS} 
All extrapolations are based on the 
PEP~II interaction region (IR) layout and bear a systematic error
of at least a factor of two. It is not trivial to predict
background levels in other IR layouts, such as that of KEK B.

The detector subsystems are subjected to different machine-related 
backgrounds. Electrons produce lost-particle backgrounds from beam-gas 
bremsstrahlung and
Coulomb scattering in addition to synchrotron radiation. Positrons just 
produce lost-particle backgrounds from beam-gas bremsstrahlung. In addition, 
the interaction of the two beams yield further backgrounds. In the 
non-colliding mode a beam-gas cross term contributes, while in colliding mode 
further backgrounds originate from luminosity and beam-beam tails. 

\section{Silicon Vertex Detectors}

Silicon vertex detectors, located closest to the beam,  
are exposed to high levels of radiation. The specific radiation dose 
depends both on beam currents and on the IR layout. For example, radiation
levels at PEP~II are expected to be higher than those at KEK~B.
In the BABAR detector the horizontal plane is most affected.
While the low-energy ring (LER) produces high radiation levels in the
forward-east region (FE-MID) of the detector, the high-energy ring 
(HER) affects 
the backward-west region (BW-MID). The dose rates in terms of beam currents 
(in units of [A]) are expected to be:

\begin{equation}
\begin{array}{rcl}
\rm
D_{SVT} \ [kRad/y]= 128 \cdot I_{LER} + 16 \cdot I^2_{LER} \ \ \ \ \ 
(FE-MID), \\
\rm
D_{SVT} \ [kRad/y]= 246 \cdot I_{HER} + 9.1 \cdot I^2_{HER} \ \  
(BW-MID).
\end{array}
\end{equation}

\begin{figure}[h]
\vskip 4.5cm
\begin{center}
\setlength{\unitlength}{1cm}
\begin{picture}(10.,5.0)
\put(-2.,-0.5)
{\mbox{\epsfysize=11.0cm\epsffile{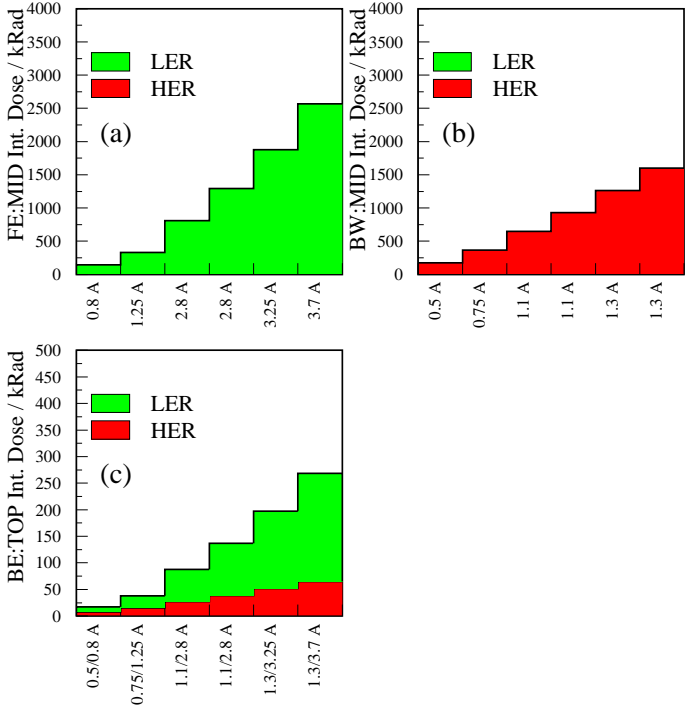}}}  
\end{picture}
\caption{Integrated-dose predictions for the BABAR SVT in the horizontal plane
versus peak beam currents, 
(a) for forward east region exposed to LER,
(b) for backward west region exposed to HER, and
(c) for top/bottom region exposed to both beams.
\label{fig:svt}}
\end{center}
\end{figure}

\noindent
For other regions the radiation dose is about a factor of ten smaller. 
Table~\ref{tab:exp} lists dose rates expected in the FE and BW regions
of the  horizontal plane for the three luminosity points. 
Figure~\ref{fig:svt} shows the total dose predictions for the BABAR 
silicon vertex tracker (SVT) in the high-radiation regions
as a function of beam currents. 
Presently, the SVT is expected to survive a 
total dose of 2~MRad. Thus, with replacement of detectors 
in the horizontal plane the BABAR SVT should 
be operational at luminosities of ${\cal L} = 1.5-3 \times 10^{34} \rm cm^{-2}
s^{-1}$.

Radiation studies of microstrip detectors in the
ATLAS semiconductor tracker (SCT) have demonstrated that silicon strip
detectors can survive very high radiation levels in hadronic 
environments, if $\rm p^+nn^+$ detectors are used that are cooled.
Before irradiation the depletion layer is 
formed between the $\rm p^+ -n$ junction. High-levels of ionization radiation
effectively change the n-type silicon to p-type.\cite{ATLAS}\cite{lutz} 
The depletion layer is shifted to 
the newly-formed $\rm p-n^+$ junction. Since for cost reasons detectors are
readout on the $\rm p^+$ side, the shifting depletion layer causes
a loss in signal. This design also introduces
a substantial amount of material in front of subsequent detectors
because of cooling and the need for twice as many layers as in a 
double-sided detector design, thus yielding increased  multiple scattering. 
Though the ATLAS design is expected to work in high-luminosity 
$e^+ e^-$ colliders, this needs to be checked in
appropriate radiation studies.

\section{Drift Chambers}

The operation of drift chambers (DCH) is affected by machine backgrounds in 
three ways. First, the total current in the drift chamber drawn by the wires
is dominated by the charge released by beam-related showers. This current 
is limited by the high-voltage system. Above this limit
the chamber becomes non-operational. Though the limit can be increased by
adding power supplies, high currents also contribute to the
aging of the drift chamber. Permanent damage is expected to occur at 
charge-densities $Q_{max} > 0.1\ \rm Cb/cm$ of wire. Second, 
the occupancy in the drift chamber due to backgrounds can hamper the 
reconstruction of physics events. Third, ionization radiation can permanently 
damage read-out electronics and digitizing electronics.

\begin{figure}[h]
\vskip 4.0cm
\begin{center}
\setlength{\unitlength}{1cm}
\begin{picture}(10.,5.0)
\put(-1.,-0.5)
{\mbox{\epsfysize=9.5cm\epsffile{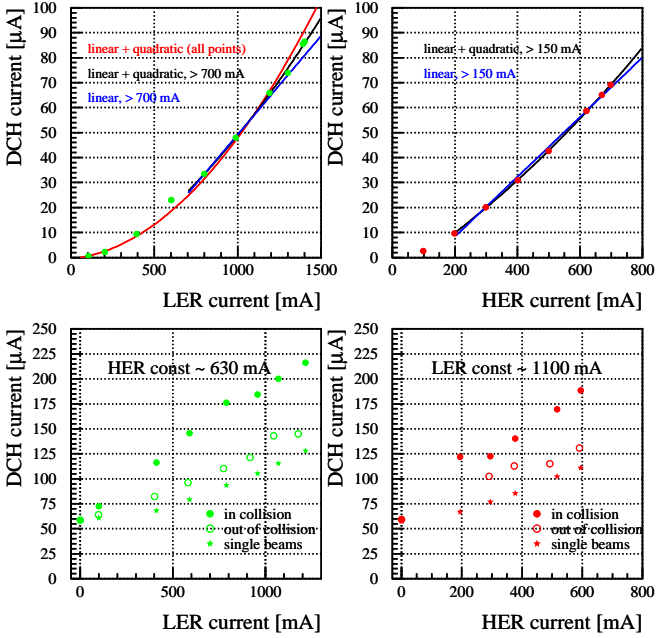}}}  
\end{picture}
\caption{BABAR DCH currents versus beam currents, top plots
for single-beam operation and bottom plots for two-beam operations.
Solid points (open circles) in bottom plots show $\rm I_{DCH}$ for
two beams in (out of) collision. The asterisk data points show the 
uncorrelated contributions from both beams. 
\label{fig:dch1}}
\end{center}
\end{figure}

\begin{figure}[h]
\vskip -.5cm
\begin{center}
\setlength{\unitlength}{1cm}
\begin{picture}(10.,5.0)
\put(1.,-0.5)
{\mbox{\epsfysize=5.5cm\epsffile{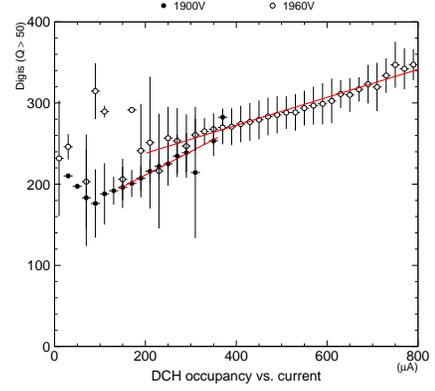}}}  
\end{picture}
\vskip 0.3cm
\caption{BABAR DCH occupancy  as a function of $\rm I_{DCH}$
for $\rm U~=~1900~V$ (solid points) and  $\rm U = 1960~V$ (open points).
\label{fig:dch2}}
\end{center}
\end{figure}

Drift chamber currents are of highest concern. Figure~\ref{fig:dch1} shows 
the total drift chamber current as functions of LER and HER currents for a 
DCH high voltage of $U = 1900~\rm V$. The data in the top left plot
were fitted with a parabolic $\rm I_{LER}$ dependence using all points and 
both linear and
parabolic $\rm I_{LER}$ dependences using points above 700~mA. 
For parameterizations in terms of $\rm I_{HER}$ (top-right plot)
both linear and linear plus
quadratic dependences for points above 150~mA were used. The best 
descriptions of the high-current regions are found for
the linear plus quadratic parameterizations which discard low-current data.
The effect of two beams in collision and out-of collision on $\rm I_{DCH}$
is shown in the bottom plots of Figure~\ref{fig:dch1}, confirming
the presence of a beam-gas cross term and luminosity-related backgrounds. 
Including all effects the total DCH current is parameterized by:
    
\begin{equation}
\begin{array}{rcl}
\rm I_{DCH}\ [\mu \rm A] = 35.3 \cdot I_{LER} + 23.5 \cdot I_{LER}^2 + 
77.2 \cdot  I_{HER} \\
\rm
+ 46.3 \cdot I_{HER}^2 + 41.9 \cdot {\cal L} -10, \ \ \ \ \ \ \ \ \ \ \ \ \ \ 
\end{array}
\end{equation}

\noindent 
where beam currents and luminosity are given in units of
[A] and [$10^{33} \rm cm^{-2} s^{-1}$], respectively. 
For $U = 1960~\rm V$, drift chamber currents scale with a factor of 1.67.  
The drift chamber occupancy at 1900 V is parameterized by:

\begin{equation}
\begin{array}{rcl}
\rm N_{DCH}\ [\%] = 0.044 + 0.191 \cdot I_{LER} + 0.0402\cdot  I_{LER}^2 
\ \ \ \ \ \ \ \ \ \  \\ 
\rm
+ 1.03 \cdot  I_{HER} + 0.113 \cdot I_{HER}^2 + 0.147 \cdot {\cal L}. 
\ \ \ \ \ \ \ \   
\end{array}
\end{equation}
\noindent
Figure~\ref{fig:dch2} shows the measured DCH occupancy in BABAR 
for high voltages of $\rm U = 1900~V$ and $\rm U = 1960~V$. 
Due to the large spread of data points an extrapolation to high 
drift chamber currents bears large uncertainties.

Table~\ref{tab:exp} lists extrapolations of the total drift chamber current,
occupancy and charge density on a wire for the three luminosity points.
At ${\cal L} = 5 \times 10^{34} \rm cm^{-2} s^{-1}$, $\rm I_{DCH}$
becomes large. For higher luminosities $\rm I_{DCH}$ basically just
scales with
${\cal L}$. In addition, an effect from the luminosity lifetime is
expected.\cite{MS}
The total DCH current definitely exceeds the tolerable limit. Thus, a 
drift chamber will not be operational and one needs to explore other 
technologies. One possibility is the use solid state devices which could be
integrated with the vertex detector. As mentioned before multiple scattering
is an issue here. But since solid state devices have much better resolutions 
than drift chambers, a reduced number of layers with thin detectors
may be acceptable. However, a detailed study needs to be devoted on this issue.

\begin{figure}[h]
\vskip 0.2cm
\begin{center}
\setlength{\unitlength}{1cm}
\begin{picture}(10.,5.0)
\put(1.5,-0.5)
{\mbox{\epsfysize=5.5cm\epsffile{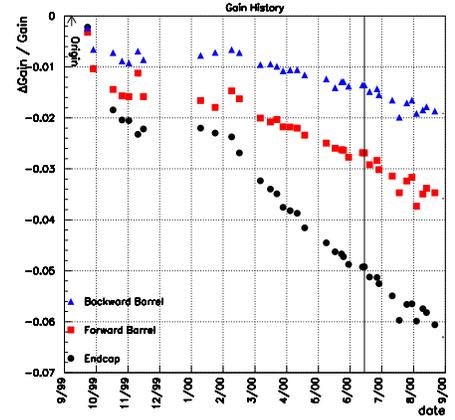}}}  
\end{picture}
\vskip 0.3cm
\caption{Gain changes of BABAR CsI(Tl) crystals in forward endcap, 
forward barrel and backward barrel versus time.
\label{fig:emcrad}}
\end{center}
\end{figure}

\begin{figure}[h]
\vskip -2.5cm
\begin{center}
\setlength{\unitlength}{1cm}
\begin{picture}(10.,5.0)
\put(-1.,-0.5)
{\mbox{\epsfysize=6.0cm\epsffile{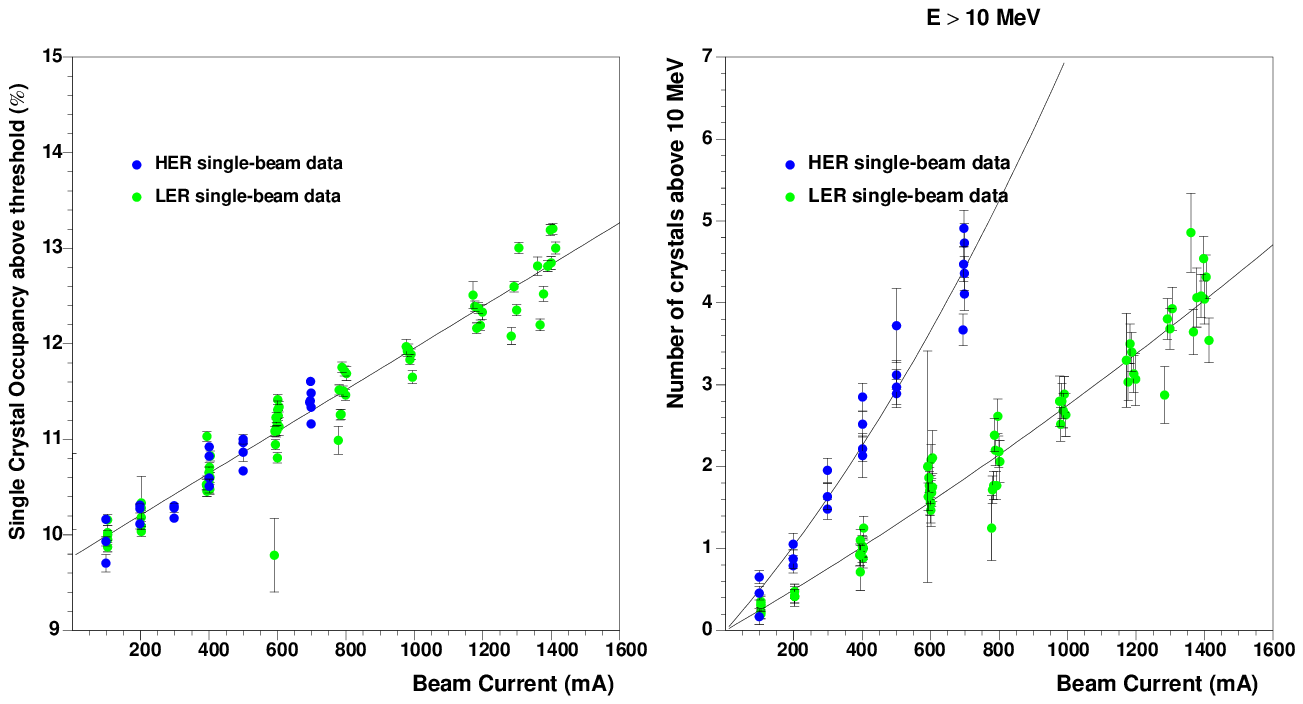}}}  
\end{picture}
\end{center}
\end{figure}

\begin{figure}[h]
\vskip -3.0cm
\begin{center}
\setlength{\unitlength}{1cm}
\begin{picture}(10.,5.0)
\put(-1.,-0.5)
{\mbox{\epsfysize=5.5cm\epsffile{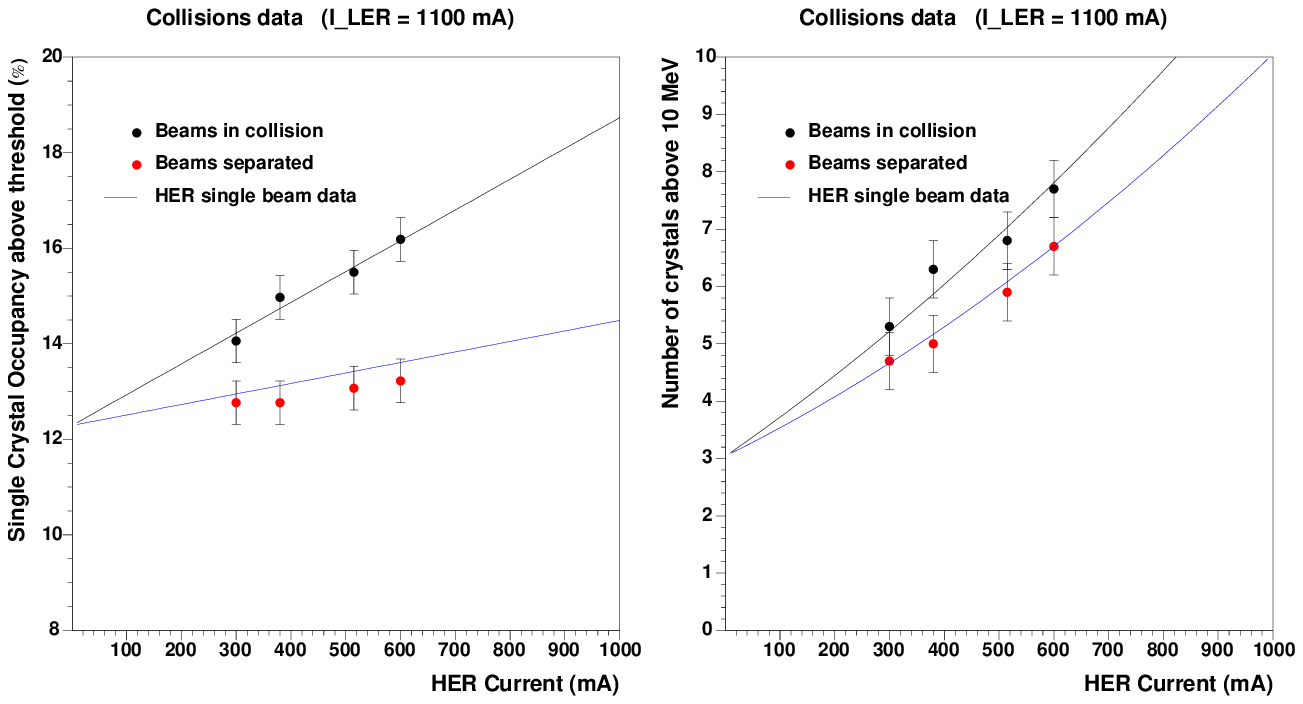}}}  
\end{picture}
\caption{Average occupancy of BABAR EMC crystals above a 1~MeV energy 
readout threshold (left) and the number of crystals containing more than 
10~MeV (right) versus beam currents. Top plots show single-beam
data and bottom plots show the $\rm I_{HER}$ dependence for fixed
$\rm I_{LER} =1100~mA$. The upper (lower) curves in bottom plots
show fits for beams in (out of) collision.
\label{fig:emc2}}
\end{center}
\end{figure}

\section{CsI(Tl) based Electromagnetic Calorimeters}

For a thallium-doped CsI crystal calorimeter radiation damage and
occupancy are the main issues. Figure~\ref{fig:emcrad} shows gain
changes in crystals located in forward and backward regions 
of the BABAR electromagnetic calorimeter (EMC). 
Forward endcap crystals, 
located closest to the beams and thus affected most by radiation,
show a gain loss of $\sim 6\%$ in the worst case. The corresponding
light-yield reductions for the worst
crystals in the forward barrel and backward barrel are 
$\sim 3.5\%$ and $\sim 2\%$, respectively.
These observations are consistent with the   
different background levels expected in these regions.
Figure~\ref{fig:emc2} shows BABAR data for the occupancy of crystals 
with energies above $\rm 1~MeV$ and for the number of crystals 
recording more than 10~MeV as a function of beam currents. 
Parameterizations in terms of beam currents ([A]) and 
luminosity ([$10^{33} \rm cm^{-2} s^{-1}$]) yield:

\begin{equation}
\begin{array}{rcl}
\rm O_{(E>1~MeV)}\ [\%] = 9.8 + 2.2 \cdot (I_{LER} + I_{HER}) 
+ 1.4 \cdot {\cal L}, \ \   \\
\rm N_{(E>10~MeV)}  = 4.7 \cdot I_{HER} + 0.23 \cdot I_{HER}^2 + 
2.4 \cdot I_{LER} \  \\
+ \ 0.33 \cdot I_{LER}^2 + 0.6 \cdot {\cal L}.\ \ \ \ \ \ \ \ \ \ \ \ \ \ \ \ 
\ \ \ 
\end{array}
\end{equation}

\noindent
Note, that for ${\cal L} > 10^{34} \rm cm^{-2} s^{-1}$ the luminosity 
contribution is expected to be dominant.
Extrapolations of $\rm O_{(E>1~MeV)}$ and $\rm N_{(E>10~MeV)}$
for the three luminosity points are presented in 
Table~\ref{tab:exp}.

For luminosities of $ {\cal L} < 1.5 \times 10^{34} \rm cm^{-2} s^{-1}$
the integrated radiation dose for CsI(Tl) crystals should be no problem, if 
observed light losses scale as expected. The impact of the large number of
low-energy photons on the EMC energy resolution needs to be studied. It
depends on the clustering algorithm and on digital filtering. 
Background rates can be reduced by improvements of the vacuum near the 
IR combined with effective collimation against $e^+$ from distant Coulomb 
scattering. For luminosities of $ {\cal L} > 1 \times 
10^{35} \rm cm^{-2} s^{-1}$, however, light losses due to radiation 
damage become substantial and high occupancy levels severely
hamper the performance.  
Therefore, present CsI(Tl) crystals are not suitable.
Other scintillators, such as pure CsI crystals, need to be 
considered and an $R \&D $ program needs to be established.

\section{Level 1 Trigger}

\begin{figure}[h]
\vskip -3.0cm
\begin{center}
\setlength{\unitlength}{1cm}
\begin{picture}(10.,5.0)
\put(-0.3,-0.5)
{\mbox{\epsfysize=6.0cm\epsffile{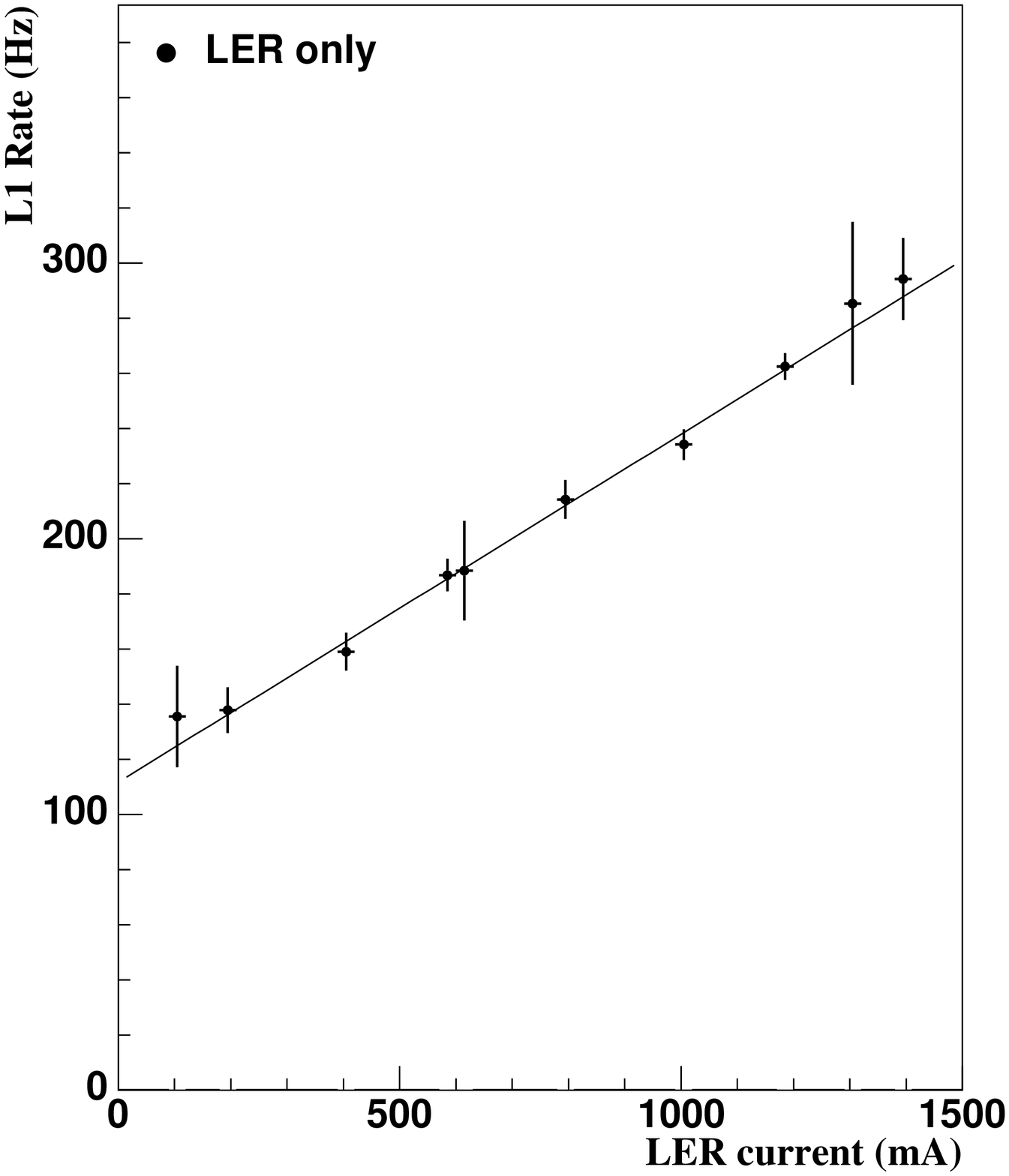}}\hspace*{-4mm}
\mbox{\epsfysize=6.0cm\epsffile{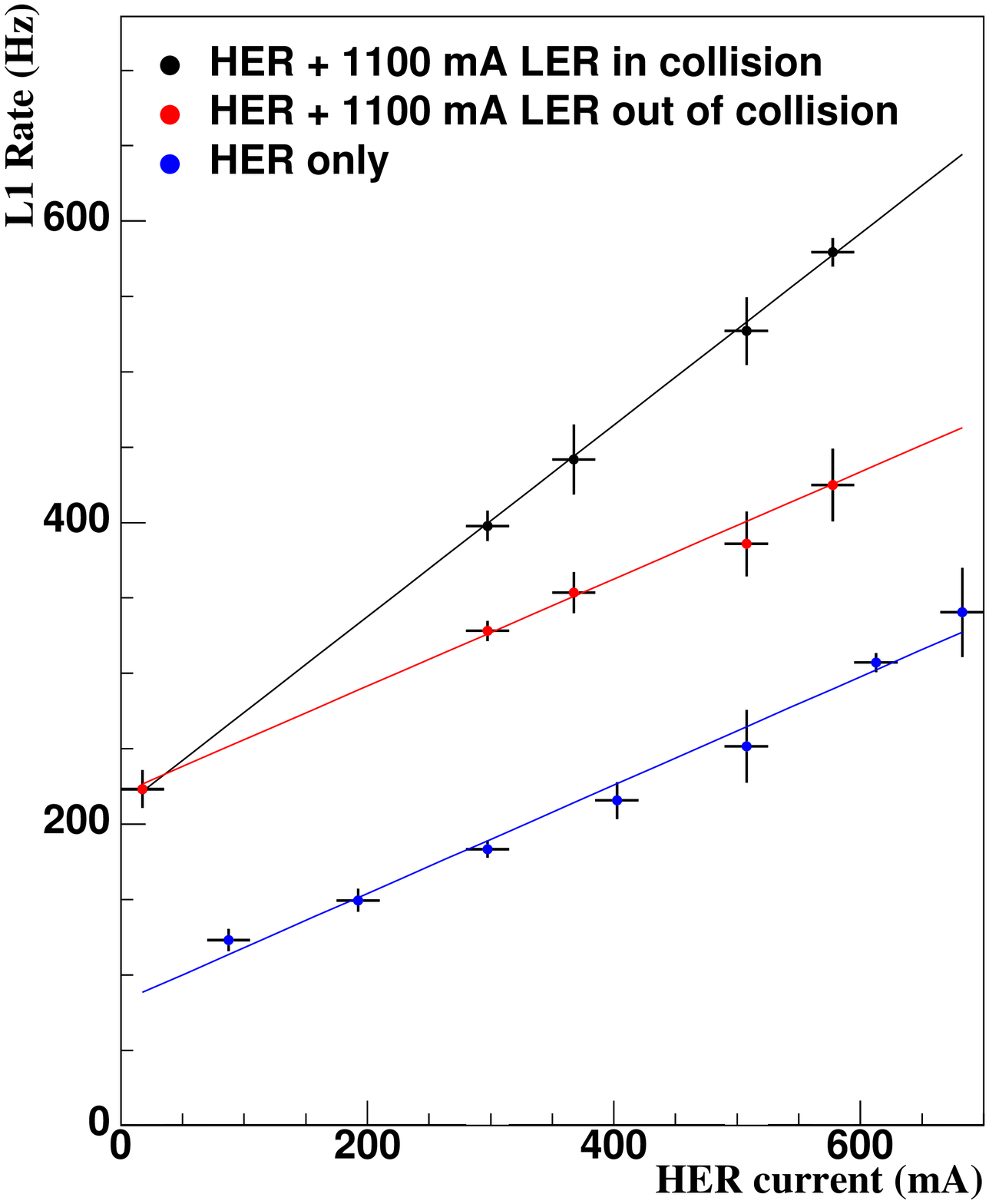}}}
\end{picture}
\caption{The L1 trigger rate versus $\rm I_{LER}$ (left) and 
$\rm I_{HER}$ (right). 
\label{fig:trig}}
\end{center}             
\end{figure}

The level 1 trigger rate scales linearly both with beam currents and with 
luminosity:

\begin{equation}
\rm 
L1\ [Hz] = 130 \ (cosmics) + 130\cdot I_{LER} + 360 \cdot I_{HER} +  
70 \cdot {\cal L}.
\end{equation}

\noindent
Extrapolations for the three luminosity points are shown
in Table~\ref{tab:exp}.
The L1 trigger rate measured in BABAR as a function of beam currents
is plotted in Figure~\ref{fig:trig}. The plot on the right
shows a comparison of single beam operation (bottom), 
two-beam operation out of collision
(middle) and two-beam operation in collision (top). 
For luminosities above $1.5 \times 10^{34} \rm cm^{-2} s^{-1}$
the BABAR trigger needs to be upgraded to cope with the increased
trigger rates. The L1 trigger rate at very high luminosities 
could be reduced by implementing a rather stringent
prescaling of Bhabhas, radiative Bhabhas and beam-gas backgrounds.
This requires an appropriate tracking device used in the trigger. 
However, LHC experiments operating at a bunch crossing of
40~MHz are designed to accept an L1 trigger rate of 100~kHz.

\begin{table}[t]
\vskip -0.3cm
\caption{Extrapolations of dose rates in silicon vertex detectors, 
total current, occupancy, and charge density in drift chambers, 
occupancy of crystals with $\rm E > 1~MeV$ and number of crystals
with $\rm E> 10~MeV$ in CsI(TL) calorimeters, 
and L1 trigger rate. 
\label{tab:exp}}
\vspace{0.2cm}
\begin{center}
\footnotesize
\begin{tabular}{|l|c|c|c|}
\hline
\raisebox{0pt}[13pt][7pt]{$ {\cal L}_{peak} \ 
\rm [cm^{-2} s^{-1}]$} &
\raisebox{0pt}[13pt][7pt]{$ 6.5 \times 10^{33}$} &
\raisebox{0pt}[13pt][7pt]{$ 1.5 \times 10^{34}$} &
\raisebox{0pt}[13pt][7pt]{$ 5.0 \times 10^{34}$} \\
\hline
\hline
\raisebox{0pt}[13pt][7pt]{$ \int {\cal L}dt \ 
\rm [fb^{-1}]$} &
\raisebox{0pt}[13pt][7pt]{$ 55$} &
\raisebox{0pt}[13pt][7pt]{$ 400$} &
\raisebox{0pt}[13pt][7pt]{$ 2000$} \\
\hline
\raisebox{0pt}[13pt][7pt]{$ \rm I_{LER}/I_{HER} \  [A]$} &
\raisebox{0pt}[13pt][7pt]{$ 2.8/1.1$} &
\raisebox{0pt}[13pt][7pt]{$ 3.7/1.3$} &
\raisebox{0pt}[13pt][7pt]{$ 4.6/1.5$} \\
\hline
\raisebox{0pt}[13pt][7pt]{$ \rm D_{SVT} \  [kRad/y]$} &
\raisebox{0pt}[13pt][7pt]{$ 480/280$} &
\raisebox{0pt}[13pt][7pt]{$ 690/340$} &
\raisebox{0pt}[13pt][7pt]{$ 1300/930$} \\
\hline
\raisebox{0pt}[13pt][7pt]{$ \rm I_{DCH} \  [\mu A]$} &
\raisebox{0pt}[13pt][7pt]{$ 680$} &
\raisebox{0pt}[13pt][7pt]{$ 1250$} &
\raisebox{0pt}[13pt][7pt]{$ 3000$} \\
\hline
\raisebox{0pt}[13pt][7pt]{$ \rm N_{DCH} \  [\%]$} &
\raisebox{0pt}[13pt][7pt]{$ 3.1$} &
\raisebox{0pt}[13pt][7pt]{$ 5$} &
\raisebox{0pt}[13pt][7pt]{$ 10$} \\
\hline
\raisebox{0pt}[13pt][7pt]{$ \rm Q_{DCH} \  [mCb/cm]$} &
\raisebox{0pt}[13pt][7pt]{$ 15$} &
\raisebox{0pt}[13pt][7pt]{$ 50$} &
\raisebox{0pt}[13pt][7pt]{$ > 100$} \\
\hline
\raisebox{0pt}[13pt][7pt]{$\rm O_{(E> 1~MeV)} \  [\%]$} &
\raisebox{0pt}[13pt][7pt]{$ 28 $} &
\raisebox{0pt}[13pt][7pt]{$ 42$} &
\raisebox{0pt}[13pt][7pt]{$ 93$} \\
\hline
\raisebox{0pt}[13pt][7pt]{$ \rm N_{(E~>~10~MeV)}  $} &
\raisebox{0pt}[13pt][7pt]{$ 21 $} &
\raisebox{0pt}[13pt][7pt]{$ 32$} &
\raisebox{0pt}[13pt][7pt]{$ 56$} \\
\hline
\raisebox{0pt}[13pt][7pt]{$ \rm L1 \ [Hz] $} &
\raisebox{0pt}[13pt][7pt]{$ 1350 $} &
\raisebox{0pt}[13pt][7pt]{$ 2130$} &
\raisebox{0pt}[13pt][7pt]{$ 4800$} \\
\hline
\end{tabular}
\end{center}
\end{table}

\section{Conclusions}

For tracking and electromagnetic energy measurements other
technologies need to be investigated. For silicon vertex detectors further
radiation studies are needed. In order to have workable solutions
for each subsystems ready, serious $\rm R \& D$ has to start soon.
A conventional muon system and a DIRC particle identification system 
with proper shielding are expected
to work at luminosities above $1 \times 10^{35} \rm cm^{-2} s^{-1}$.

\section*{Acknowledgments}

I would like to thank W. Kozanecki D. MacFarlane and B. Stugu for useful 
discussions. This work has been supported by the Research Council
of Norway.

\end{document}